\documentclass[sigplan]{acmart}

\AtBeginDocument{%
  }

\usepackage{enumitem}

\providecommand{\eg}{\textit{e.g.}}
\providecommand{\etal}{\textit{et al.}}

\setcopyright{acmlicensed}
\copyrightyear{2026}
\acmYear{2026}
\acmDOI{10.48550/arXiv.2602.12430}
\acmConference[AgentSkills '26]{ACM CAIS 2026 Workshop AgentSkills}{May 26, 2026}{San Jose, USA}

\begin{document}

\title{Agent Skills for Large Language Models: Architecture,
  Acquisition, Security, and the Path Forward}

\author{Renjun Xu}
\affiliation{%
  \institution{ReDiscovery}
  \city{Hangzhou}
  \country{China}
}
\email{rux@zju.edu.cn}

\author{Yang Yan}
\affiliation{%
  \institution{Westlake University}
  \city{Hangzhou}
  \country{China}
}
\email{yanyang@westlake.edu.cn}

\renewcommand{\shortauthors}{Xu and Yan}

\begin{abstract}
The transition from monolithic language models to modular, skill-equipped agents marks a defining shift in how large language models (LLMs) are deployed in practice.
Rather than encoding all procedural knowledge within model weights, agent skills---composable packages of instructions, code, and resources that agents load on demand---enable dynamic capability extension without retraining. It is formalized in a paradigm of progressive disclosure, portable skill definitions, and integration with the Model Context Protocol (MCP).
This survey provides a comprehensive treatment of the agent skills landscape, as it has rapidly evolved during the last few months.
We organize the field along four axes: (i) architectural foundations, examining the SKILL.md specification, progressive context loading, and the complementary roles of skills and MCP; (ii) skill acquisition, covering reinforcement learning with skill libraries (SAGE), autonomous skill discovery (SEAgent), and compositional skill synthesis; (iii) deployment at scale, emphasizing computer-use agents (CUAs) while situating them alongside software-engineering, web, tool-use, and embodied-agent benchmarks; and (iv) security, where recent empirical analyses report that 26.1\% of community-contributed skills contain vulnerabilities, motivating our proposed Skill Trust and Lifecycle Governance Framework---a four-tier, gate-based permission model that maps skill provenance to graduated deployment capabilities.
We identify seven open challenges---from cross-platform skill portability to capability-based permission models---and propose a research agenda for realizing trustworthy, self-improving skill ecosystems.
Unlike prior surveys that broadly cover LLM agents or tool use, this work focuses specifically on the emerging skill abstraction layer and its implications for the next generation of agentic systems.
An accompanying collection of agent skills resources is available at \url{https://github.com/scienceaix/agentskills}.
\end{abstract}



\maketitle

\section{Introduction}
\label{sec:intro}

The capabilities of large language models have expanded dramatically, yet their utility as autonomous agents remains constrained by a fundamental tension: general-purpose models possess broad knowledge but lack the specialized procedural expertise that real-world tasks demand. Fine-tuning addresses this partially, but at significant cost and with limited composability. Retrieval-augmented generation (RAG) provides external knowledge, but the retrieved passages are passive---they cannot prescribe multi-step workflows, bundle executable code, or adapt the agent's tool permissions at runtime.

\textit{Agent skills} resolve this tension by introducing a modular, filesystem-based abstraction that equips agents with domain-specific expertise on demand. In this paradigm, a skill is not a model or a prompt template, but a self-contained package: a structured instruction file (SKILL.md), optional scripts, reference documents, and assets, organized in a directory that the agent discovers, loads, and follows when relevant tasks arise~\cite{zhang2025skills}. The distinction from traditional tools is architectural: tools execute and return results, whereas skills \textit{prepare} the agent to solve a problem by injecting procedural knowledge, modifying execution context, and enabling progressive disclosure of information.

Anthropic formalized this concept in October 2025 with the launch of Agent Skills across Claude's product surface~\cite{anthropic2025skillslaunch}, followed by its release as an open standard in December 2025~\cite{agentskillsio}. Within four months, the \texttt{anthropics/skills} repository accumulated over 62,000 GitHub stars, partner-built skills from Atlassian, Figma, Canva, Stripe, and Notion entered a curated directory, and structurally identical architectures were independently adopted by other frontier model providers.

This rapid convergence reflects a broader recognition: as agents move from research prototypes to production deployments, the industry needs standardized mechanisms for packaging, distributing, and governing procedural expertise. The parallel maturation of the Model Context Protocol (MCP)---an open standard for connecting agents to external data and tools, donated to the Linux Foundation's Agentic AI Foundation in December 2025~\cite{anthropic2025mcpaaif}---provides the complementary infrastructure layer. Together, skills and MCP define an emerging \textit{agentic stack} in which skills supply the ``what to do'' and MCP supplies the ``how to connect.''

This survey is the first to provide a focused treatment of the agent skills paradigm. While excellent surveys exist on LLM agents broadly~\cite{luo2025llmagentsurvey,plaat2025agentic}, tool use~\cite{qu2025toollearning}, and GUI agents~\cite{zhang2024guisurvey,hu2025osagents}, none specifically examines the \textit{skill abstraction layer} that has emerged as a unifying architectural primitive. Our contributions are:

\begin{enumerate}[leftmargin=*,topsep=4pt,itemsep=2pt]
    \item A systematic analysis of the Agent Skills architecture, including progressive disclosure, the SKILL.md specification, and its relationship to MCP (Section~\ref{sec:architecture}).
    \item A taxonomy of skill acquisition methods spanning reinforcement learning, autonomous exploration, compositional synthesis, and human-authored knowledge packaging (Section~\ref{sec:acquisition}).
    \item A critical assessment of deployment settings for skills, emphasizing the computer-use agent (CUA) stack while situating it within broader software-engineering, web, function-calling, and embodied-agent evaluation contexts (Section~\ref{sec:deployment}).
    \item The first consolidated treatment of agent skill security, synthesizing three concurrent empirical studies, and an original \textbf{Skill Trust and Lifecycle Governance Framework} that maps acquisition provenance through verification gates to graduated deployment permissions (Section~\ref{sec:security}).
    \item An articulation of seven open challenges and a research agenda for the field (Section~\ref{sec:challenges}).
\end{enumerate}

\section{Background and Scope}
\label{sec:background}

\subsection{From Prompt Engineering to Skill Engineering}

The evolution toward agent skills can be understood as a progression through three paradigms of LLM capability extension. \textit{Prompt engineering} (2022--2023) demonstrated that carefully crafted instructions could elicit impressive zero-shot and few-shot behaviors, but prompts are ephemeral, non-modular, and difficult to version or share. \textit{Tool use and function calling} (2023--2024) enabled models to invoke external APIs, but each tool is atomic---a single function with defined inputs and outputs. Tools execute and return; they do not reshape the agent's understanding of a task.

\textit{Skill engineering} (2025--present) introduces a higher-order abstraction. A skill is a \textit{bundle} that can include instructions, workflow guidance, executable scripts, reference documentation, and metadata, all organized to be dynamically loaded when relevant. The key insight is that many real-world tasks require not a single tool call but a coordinated sequence of decisions informed by domain-specific procedural knowledge. A PDF-processing skill, for example, does not merely expose a ``fill form'' function; it teaches the agent \textit{how} to approach PDF manipulation, which libraries to use, what edge cases to handle, and what code to execute~\cite{zhang2025skills}.

\subsection{Relationship to Prior Work}

Several foundational works anticipated the skill paradigm. Voyager~\cite{wang2023voyager} introduced a skill library for embodied agents in Minecraft, where LLM-generated programs were stored and composed to solve increasingly complex tasks. CREATOR~\cite{qian2024creator} and Large Language Models as Tool Makers~\cite{cai2024large} explored the idea that LLMs could create their own tools. Toolformer~\cite{schick2024toolformer} demonstrated self-taught tool use. However, these works focused on \textit{model-generated} skills in constrained environments. The Agent Skills paradigm, by contrast, emphasizes \textit{human-authored, portable, and governed} skill packages designed for production deployment across heterogeneous agent platforms.

\subsection{Scope and Methodology}

In this survey, we systematically searched arXiv, ACL Anthology, NeurIPS/ICML/ICLR proceedings, and official technical publications using queries centered on ``agent skills,'' ``skill library,'' ``LLM tool use,'' ``computer-use agent,'' and ``Model Context Protocol.'' We explicitly exclude earlier surveys' coverage of general LLM agent architectures and broad tool-use taxonomies, focusing instead on the skill abstraction layer and its immediate ecosystem. Because skill packaging has first matured in software-facing agent environments, our deployment section gives CUAs extended treatment; nevertheless, we treat this as a representative stress case rather than the full boundary of the paradigm.

\section{Architectural Foundations}
\label{sec:architecture}

\subsection{The SKILL.md Specification}

At its core, a skill is a directory containing a SKILL.md file with YAML frontmatter that specifies a \texttt{name} and \texttt{description}. The agent pre-loads only this metadata---typically a few dozen tokens---into its system prompt at startup, enabling large skill libraries without context penalty. The full body of the SKILL.md file contains procedural instructions that are loaded only when the skill is triggered. Additional resources (scripts, reference docs, assets) reside in subdirectories and are loaded on demand~\cite{zhang2025skills}.

\begin{figure}[htbp]
\centering
\includegraphics[width=\columnwidth]{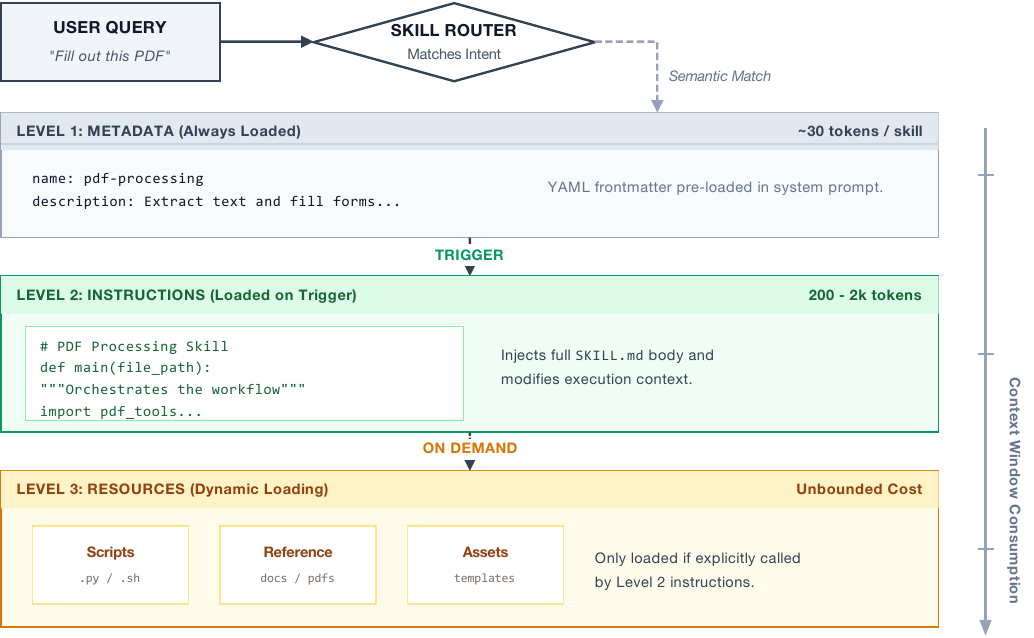}
\caption{Progressive disclosure architecture of Agent Skills. Information is loaded in three stages to minimize context window consumption while maintaining access to arbitrarily deep procedural knowledge. Token estimates are approximate per-skill averages; adapted from Zhang, Lazuka, and Murag~\cite{zhang2025skills}.}
\Description{A three-tier diagram showing progressive disclosure of agent skills, with metadata at the top, full SKILL.md body in the middle, and bundled scripts and reference documents at the bottom.}
\label{fig:progressive}
\end{figure}

This three-level progressive disclosure, summarized in Figure~\ref{fig:progressive}, is the defining architectural innovation. As Zhang \etal{} articulate: building a skill is ``like putting together an onboarding guide for a new hire''~\cite{zhang2025skills}. Level~1 serves as a table of contents; Level~2 provides the chapter content; Level~3 supplies the technical appendices.

\subsection{Skill Execution Lifecycle}

When a user request matches a skill's description, the agent triggers a two-phase execution. First, the skill's instructions and any required resources are injected into the conversation context as a hidden (meta) message---visible to the model but not rendered in the user interface. Second, the agent's execution context is modified: pre-approved tools (\eg, specific bash commands, file read/write permissions) are activated, and the agent proceeds with the enriched context to complete the task.

Critically, skill execution modifies the \textit{agent's preparation}, not its output directly. This distinguishes skills from function calls, where the tool produces a result. A skill reshapes what the agent knows and can do before it generates its response.

\subsection{The Agentic Stack: Skills and MCP}

The Model Context Protocol (MCP), launched in November 2024~\cite{anthropic2024mcp} and donated to the Agentic AI Foundation in December 2025~\cite{anthropic2025mcpaaif}, provides a complementary layer. MCP standardizes how agents connect to external data sources and tools via a JSON-RPC 2.0 protocol with three primitives: \textit{tools} (model-invoked functions), \textit{resources} (application-controlled data), and \textit{prompts} (user-invoked templates)~\cite{mcpspec2025}.

Skills and MCP are not competing standards but orthogonal layers of an emerging agentic stack (Table~\ref{tab:skills_vs_mcp}). A skill might instruct the agent to use a particular MCP server, specify how to interpret its outputs, and define fallback strategies if the connection fails. Skills provide the procedural intelligence; MCP provides the connectivity.

\begin{table*}[t]
\centering
\caption{Comparison of Agent Skills and MCP as complementary layers of the agentic stack.}
\label{tab:skills_vs_mcp}
\small
\begin{tabular}{@{}p{2.5cm} p{6.0cm} p{6.0cm}@{}}
\toprule
\textbf{Dimension} & \textbf{Agent Skills} & \textbf{MCP} \\
\midrule
Primary role & Procedural knowledge & Tool connectivity \\
Unit & Directory w/ SKILL.md & Server w/ endpoints \\
Loaded by & Agent on trigger & Client on config \\
Modifies & Context + permissions & Available tools/data \\
Persistence & Filesystem-based & Session-based \\
Spec status & Open std (Dec 2025) & Linux Found. (Dec 2025) \\
\bottomrule
\end{tabular}
\end{table*}

\subsection{Advanced Tool Use Integration}

Anthropic's November 2025 release of Advanced Tool Use features~\cite{anthropic2025advancedtools} introduced three mechanisms that deepen the skill--tool integration: (1) a \textit{Tool Search Tool} that enables programmatic discovery of relevant tools from large registries, with the cited technical report reporting token-overhead reductions of up to 85\% relative to pre-loading tool definitions; (2) \textit{Programmatic Tool Calling}, where the model executes tools via code rather than structured JSON, with reported accuracy improving from 79.5\% to 88.1\% on Opus 4.5 in the reported benchmark; and (3) \textit{Tool Learning}, which allows the model to study tool documentation and improve invocation quality. These features address a practical bottleneck: as skill libraries grow, the agent needs efficient mechanisms to discover and invoke the right tools within a skill's workflow.

\section{Skill Acquisition and Learning}
\label{sec:acquisition}

A central question in the skills paradigm is how skills are created, refined, and composed. We identify four distinct acquisition modalities, summarized in Table~\ref{tab:acquisition}.

\begin{table*}[t]
\centering
\caption{Taxonomy of skill acquisition methods for LLM agents (2025--2026). Performance metrics are from the original papers on their respective benchmarks.}
\label{tab:acquisition}
\small
\begin{tabular}{@{}p{2.6cm}p{3.2cm}p{2.6cm}p{2.8cm}p{3.0cm}@{}}
\toprule
\textbf{Method} & \textbf{Representative Work} & \textbf{Mechanism} & \textbf{Benchmark} & \textbf{Key Result} \\
\midrule
Human-authored & Anthropic Skills~\cite{zhang2025skills} & Manual SKILL.md & Claude products & 62k+ GitHub stars \\
RL with skill library & SAGE~\cite{wang2025sage} & GRPO + sequential rollout & AppWorld & +8.9\% SGC, $-$59\% tokens \\
Autonomous exploration & SEAgent~\cite{sun2025seagent} & Curriculum + world model & OSWorld (5 apps) & 11.3\%$\to$34.5\% success \\
Structured skill base & CUA-Skill~\cite{chen2026cuaskill} & Parameterized exec. graphs & WindowsAgentArena & 57.5\% SOTA \\
Compositional synthesis & Agentic Proposing~\cite{jiao2026agenticproposing} & Skill library + GoT agent & AIME 2025 & 91.6\% (30B solver) \\
Skill compilation & Li 2026~\cite{li2026singleagent} & Multi-agent $\to$ single-agent & Multiple & Phase transition found \\
\bottomrule
\end{tabular}
\end{table*}

\subsection{Human-Authored Skills}

The most immediately impactful acquisition modality is direct human authoring. Anthropic's framework is deliberately designed to make this accessible: a skill can be as simple as a Markdown file with a few dozen lines of instructions~\cite{zhang2025skills}. The \texttt{skill-creator} meta-skill within Claude Code can scaffold a new skill from a natural-language description, generating the directory structure, SKILL.md, and bundled scripts. Enterprise deployments at companies such as Atlassian, Canva, and Sentry have produced production-grade skills that encode proprietary workflows.

The December 2025 partner directory launch~\cite{anthropic2025skillslaunch} established a curation pipeline: partners submit skills that are reviewed for security and quality before inclusion. This model mirrors app store governance but with significantly lower barrier to entry, since skills are essentially structured documents rather than executable applications.

\subsection{Reinforcement Learning with Skill Libraries}

SAGE (Skill Augmented GRPO for self-Evolution)~\cite{wang2025sage} represents the most rigorous approach to learning skills through reinforcement. The key innovation is \textit{Sequential Rollout}: rather than training on isolated tasks, the agent is deployed across chains of similar tasks, with skills generated in earlier tasks preserved and available for reuse in subsequent ones. A \textit{Skill-integrated Reward} combines outcome-based verification with an additional signal rewarding high-quality, reusable skill creation.

On AppWorld, SAGE achieves 72.0\% Task Goal Completion and 60.7\% Scenario Goal Completion---an 8.9\% absolute improvement over baseline GRPO without skill libraries---while requiring 26\% fewer interaction steps and 59\% fewer generated tokens. This efficiency gain is particularly significant for production deployment, where token consumption directly translates to cost.

\subsection{Autonomous Skill Discovery}

SEAgent~\cite{sun2025seagent} addresses a complementary challenge: can agents autonomously discover skills for previously unseen software? The framework introduces a \textit{World State Model} for stepwise trajectory assessment and a \textit{Curriculum Generator} that produces increasingly complex tasks from a continuously updated software guidebook memory. A specialist-to-generalist training strategy integrates insights from domain-specific agents into a unified model. On five novel software environments in OSWorld, SEAgent improves success rates from 11.3\% to 34.5\%---a 23.2 percentage point gain over the competitive UI-TARS baseline.

\subsection{Structured Skill Bases}

CUA-Skill~\cite{chen2026cuaskill} takes a knowledge-engineering approach, encoding human computer-use expertise as parameterized \textit{execution graphs} and \textit{composition graphs}. Each skill has typed parameters, preconditions, and composability rules. The CUA-Skill Agent supports dynamic skill retrieval, argument instantiation, and memory-aware failure recovery. On WindowsAgentArena, this achieves a state-of-the-art 57.5\% success rate, substantially outperforming approaches that lack structured skill representations.

\subsection{Compositional Skill Synthesis}

Agentic Proposing~\cite{jiao2026agenticproposing} demonstrates that skills can be composed dynamically during problem-solving. A specialized agent selects and composes modular reasoning skills from a library, modeling problem synthesis as a goal-driven process. The paper reports that a 30B parameter solver using this approach achieves 91.6\% on the AIME 2025 mathematical competition benchmark; within the scope of the cited evaluation, this result illustrates how skill composition can yield capabilities that exceed what any individual skill provides.

\subsection{Skill Compilation: Multi-Agent to Single-Agent}

A provocative finding from Li~\cite{li2026singleagent} is that multi-agent systems can often be ``compiled'' into single-agent skill libraries with substantial reductions in token usage and latency while maintaining accuracy. However, this compression exhibits a \textit{phase transition}: beyond a critical library size, skill selection accuracy degrades sharply. This finding has practical implications for skill library management: there exist fundamental limits on how many skills a single agent can effectively manage.

\section{Deployment: Agents, Benchmarks, and the Computer-Use Stack}
\label{sec:deployment}

Computer-use agents (CUAs) are the most visible deployment domain for the skills paradigm, since operating a computer through a GUI inherently requires composing sequences of perception, reasoning, and action that map naturally to skill abstractions. They are not the only relevant domain: software-engineering agents, web agents, function-calling systems, embodied agents, scientific assistants, and educational tutors all require reusable procedural knowledge. CUAs receive extended treatment here because their benchmarks expose the full perception--grounding--action loop and because their failure modes are unusually concrete.

\subsection{GUI Agent Architectures}

\begin{figure}[htbp]
\centering
\includegraphics[width=\columnwidth]{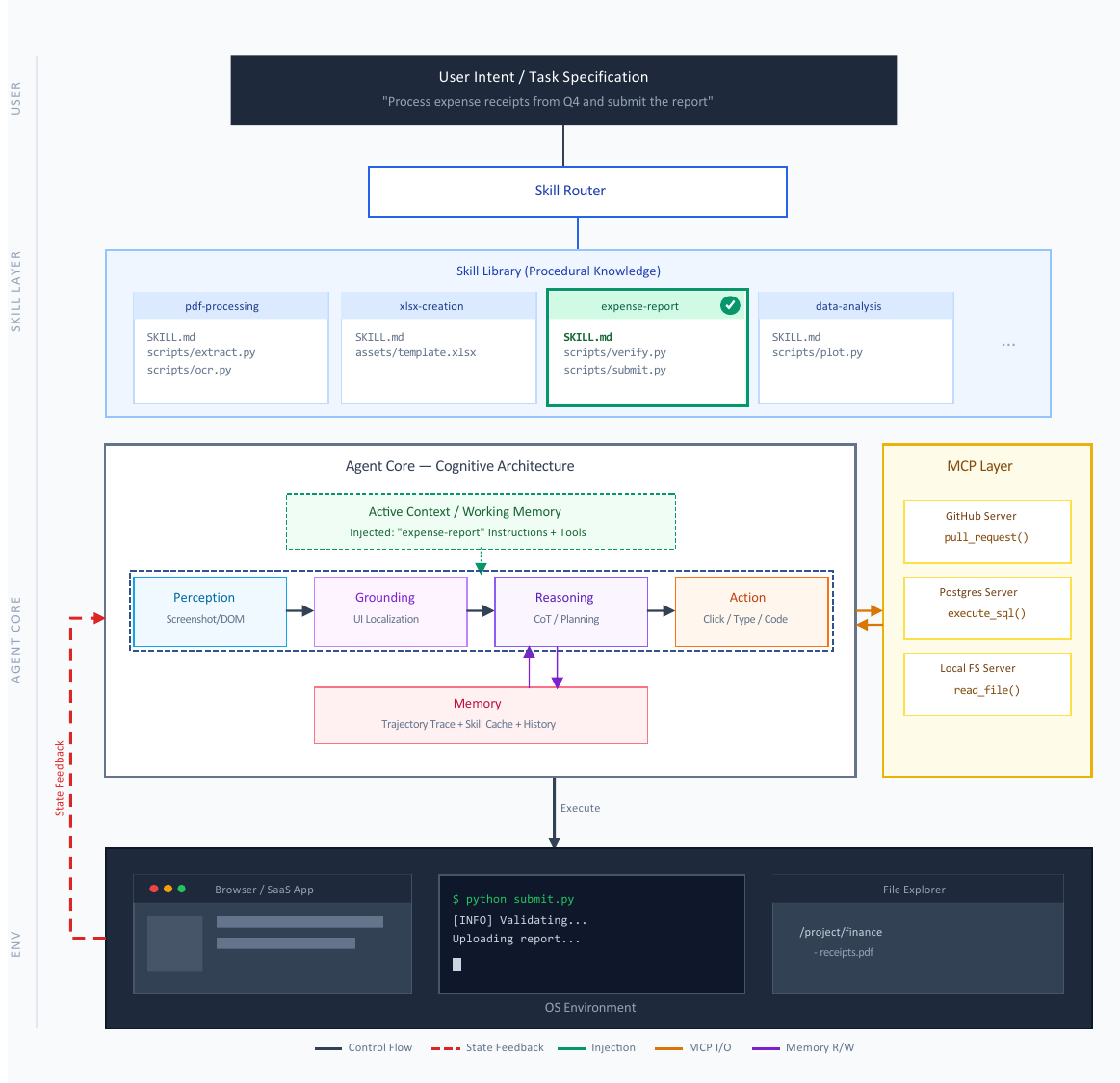}
\caption{Architecture of a skill-equipped computer-use agent showing the interplay between the skill library, perception-grounding-action pipeline, MCP connectivity layer, and the operating system environment. The active skill (highlighted) is selected by the router and injected into the agent's context.}
\Description{A block diagram of a computer-use agent showing a skill library feeding into a perception-grounding-action pipeline, with an MCP layer connecting to the operating system.}
\label{fig:cua}
\end{figure}

Figure~\ref{fig:cua} shows how a skill-equipped CUA composes a skill library, perception and grounding modules, MCP connectivity, and operating-system actions. The state of the art in CUA architectures has progressed rapidly. UI-TARS~\cite{qin2025uitars}, introduced in January 2025, established new baselines across ten GUI benchmarks through enhanced perception, unified action modeling, and System-2 reasoning. Its successor, UI-TARS-2~\cite{wang2025uitars2}, introduced a data flywheel for scalable trajectory generation and stabilized multi-turn RL training, reaching 47.5\% on OSWorld and 73.3\% on AndroidWorld.

Agent S2~\cite{agashe2025agents2} proposed a compositional generalist-specialist framework with a Mixture-of-Grounding mechanism for precise GUI localization, achieving 18.9\% and 32.7\% relative improvements over Claude Computer Use and UI-TARS, respectively, on OSWorld.

OpenCUA~\cite{wang2025opencua} provides the most comprehensive open-source framework, with AgentNet---the first large-scale CUA dataset spanning three operating systems and over 200 applications. OpenCUA-72B achieves 45.0\% on OSWorld-Verified, the strongest open-source result. Its release at NeurIPS 2025 as a Spotlight paper signals the field's maturity.

\subsection{GUI Grounding Advances}

Accurate GUI grounding---identifying the correct screen element to interact with---remains a critical skill for CUAs. UGround~\cite{gou2025uground}, an ICLR 2025 Oral, trained on 10 million GUI elements from 1.3 million screenshots, establishing the largest GUI visual grounding dataset and outperforming existing models by up to 20\% absolute. The Jedi framework~\cite{xie2025jedi} scaled grounding data to 4 million examples via UI decomposition and synthesis, improving OSWorld agentic success from 5\% to 27\%.

A striking result from Yuan \etal{}~\cite{yuan2025visualgrounding} demonstrates that RL-based self-evolutionary training enables a 7B parameter model to achieve 47.3\% on ScreenSpot-Pro, outperforming the 72B UI-TARS model by 24.2 percentage points with only 3,000 training samples. Similarly, GUI-Actor~\cite{wu2025guiactor} introduces coordinate-free visual grounding via an attention-based action head, with a 7B model surpassing UI-TARS-72B on ScreenSpot-Pro.

\subsection{Benchmark Landscape}

Table~\ref{tab:benchmarks} presents representative benchmark results for CUA-adjacent and tool-use settings. The field has witnessed rapid progress: on OSWorld-style tasks, reported success rates have climbed from single digits in early 2024 to benchmark-specific parity with the published human baseline in OSWorld-V (72.6\% vs.\ 72.36\%) by December 2025. This should not be read as generalized operational superiority over human users; rather, it indicates that narrow benchmark slices are beginning to saturate. More challenging settings---professional applications (ScreenSpot-Pro), long-horizon tasks (OS-Marathon~\cite{osmarathon2026}), hybrid GUI-code workflows (CoAct-1~\cite{coact2025}), software maintenance tasks (SWE-bench~\cite{jimenez2024swebench}), and function-calling evaluations (BFCL~\cite{gorilla2025bfcl})---continue to expose significant gaps.

\begin{table}[t]
\centering
\caption{Representative agent benchmark results (as of Feb 2026). SR = Success Rate (\%).}
\label{tab:benchmarks}
\small
\begin{tabular}{@{}lllr@{}}
\toprule
\textbf{Benchmark} & \textbf{Reported Agent} & \textbf{SR} & \textbf{Human} \\
\midrule
OSWorld & CoAct-1 & 59.9 & 72.4 \\
OSWorld-V & Proprietary & 72.6 & 72.4 \\
WinAgentArena & CUA-Skill & 57.5 & -- \\
AndroidWorld & UI-TARS-2 & 73.3 & -- \\
ScreenSpot-Pro & Yuan \etal & 47.3 & -- \\
SWE-bench V. & Claude Opus 4.6 & 79.2 & -- \\
BFCL v4 & Multiple & $>$90 & -- \\
\bottomrule
\end{tabular}
\end{table}

\section{Security of Agent Skills}
\label{sec:security}

The rapid adoption of agent skills has introduced a significant and novel attack surface. Unlike traditional software packages, skills combine natural-language instructions with executable code in a format that agents trust implicitly. Three concurrent studies, all appearing between October 2025 and February 2026, provide the first empirical characterization of this threat landscape.

\subsection{Prompt Injection via Skills}

Schmotz \etal{}~\cite{schmotz2025skillinjection} demonstrated that Agent Skills enable ``trivially simple'' prompt injections. By embedding malicious instructions within long SKILL.md files and referenced scripts, attackers can exfiltrate sensitive data such as internal files or passwords. Critically, the authors show that system-level guardrails of a popular coding agent can be bypassed: a benign, task-specific approval with a ``Don't ask again'' option carries over to closely related but harmful actions. The attack exploits the fundamental trust model of skills---once loaded, a skill's instructions are treated as authoritative context.

\subsection{Vulnerabilities at Scale}

Liu \etal{}~\cite{liu2026skillswild} conducted the first large-scale empirical security analysis, collecting 42,447 skills from two major marketplaces and analyzing 31,132 using SkillScan, a multi-stage detection framework combining static analysis with LLM-based semantic classification. The findings are sobering: \textbf{26.1\% of skills contain at least one vulnerability}, spanning 14 distinct patterns across four categories: prompt injection, data exfiltration (13.3\%), privilege escalation (11.8\%), and supply chain risks. Skills bundling executable scripts are 2.12$\times$ more likely to contain vulnerabilities than instruction-only skills (OR=2.12, $p<0.001$). 5.2\% of skills exhibit high-severity patterns strongly suggesting malicious intent.

\subsection{Confirmed Malicious Skills}

A subsequent study~\cite{liu2026maliciousskills} constructed the first ground-truth dataset of confirmed malicious skills by behaviorally verifying 98,380 skills from two community registries. Among 157 confirmed malicious skills with 632 vulnerabilities, the authors identified two attack archetypes: \textit{Data Thieves} that exfiltrate credentials through supply chain techniques, and \textit{Agent Hijackers} that subvert agent decision-making through instruction manipulation. A single industrialized actor accounted for 54.1\% of confirmed cases through templated brand impersonation.

\begin{figure*}[htbp]
\centering
\includegraphics[width=\textwidth]{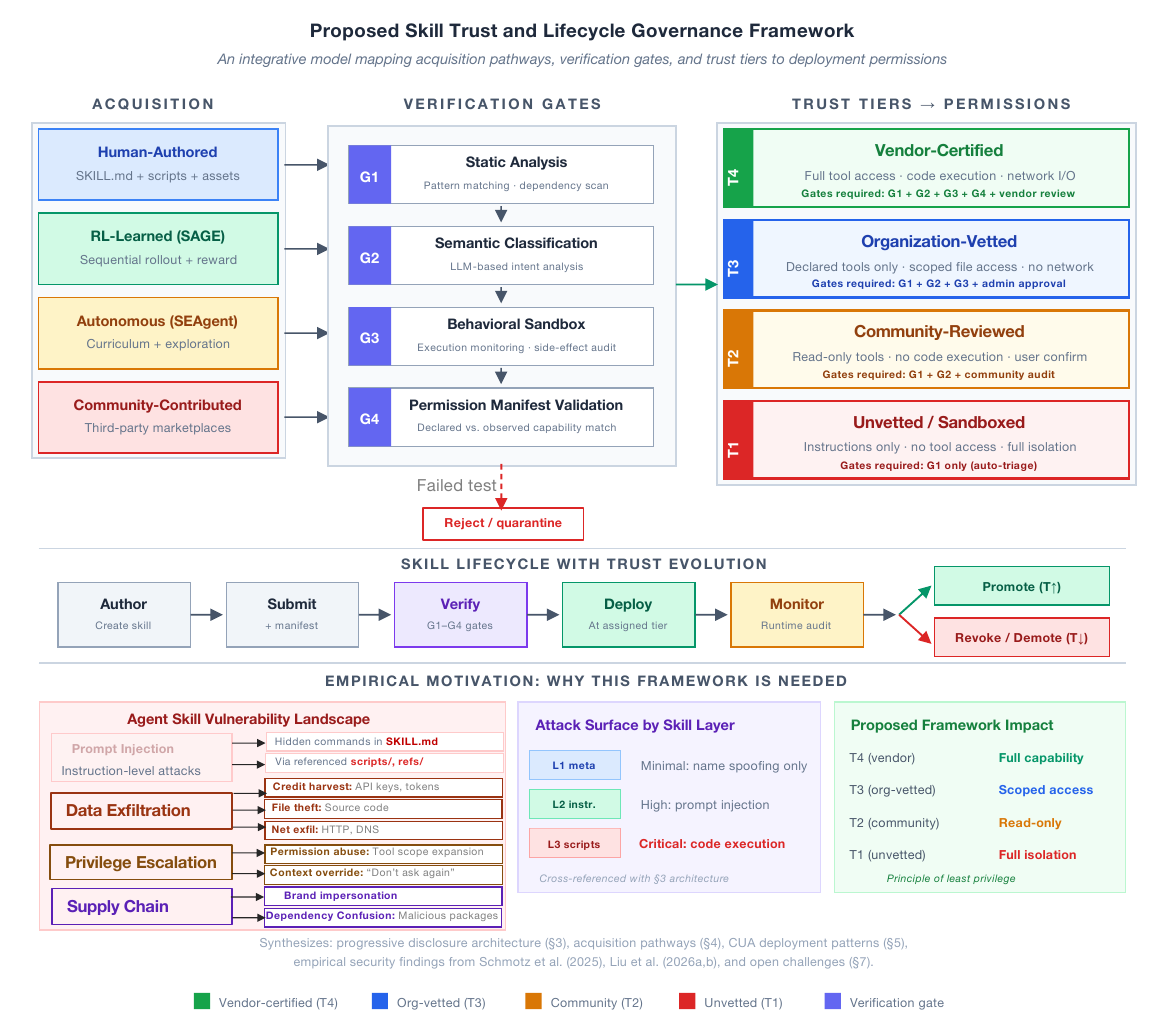}
\caption{Skill Trust and Lifecycle Governance Framework: acquisition pathways pass through verification gates, map to graduated trust tiers, and evolve through runtime monitoring.}
\Description{A framework diagram showing skill acquisition pathways on the left, feeding through four sequential verification gates G1 through G4, and resulting in four graduated trust tiers T1 through T4 with escalating deployment permissions, with a lifecycle feedback loop at the bottom.}
\label{fig:trust}
\end{figure*}

\subsection{Toward a Governance Framework}

The findings above---from prompt injection via trusted skill files~\cite{schmotz2025skillinjection} to the 26.1\% vulnerability rate across 42,447 community skills~\cite{liu2026skillswild} to confirmed malicious actors operating at scale~\cite{liu2026maliciousskills}---collectively demonstrate that the current implicit trust model is untenable. The governance problem is not only whether a skill is malicious at installation time, but also whether its declared purpose, embedded instructions, scripts, dependencies, and runtime behavior remain aligned as the environment changes.

We propose a \textbf{Skill Trust and Lifecycle Governance Framework} (Figure~\ref{fig:trust}) that synthesizes insights from across this survey into a principled governance model. The figure's attack categories summarize the concrete families reported in the empirical studies, while the gates and tiers express a deployment policy for reducing their impact. The framework has three components:

\textit{(1) Verification gates.} Four sequential gates (G1--G4) provide defense in depth. G1 applies static analysis---pattern matching and dependency scanning---to flag known vulnerability signatures. G2 uses LLM-based semantic classification to detect intent mismatches between a skill's declared purpose and its actual instructions, addressing the indirect prompt injection vectors identified by Schmotz \etal{}~\cite{schmotz2025skillinjection}. G3 executes the skill in a behavioral sandbox to detect side effects invisible to static analysis, motivated by the finding that confirmed malicious skills average 4.03 vulnerabilities spanning 3 kill-chain phases~\cite{liu2026maliciousskills}. G4 validates a proposed \textit{permission manifest}: a formal declaration of required capabilities (tools, file paths, network access) that is compared against observed behavior from G3.

\textit{(2) Trust tiers.} The framework assigns skills to one of four tiers (T1--T4) based on the gates they pass and their provenance. The key insight is that the mapping is not binary (safe/unsafe) but graduated, following the principle of least privilege. An unvetted community skill (T1) receives instructions-only access with full tool isolation. A vendor-certified skill (T4) receives full capabilities. This directly addresses the finding that bundling executable scripts increases vulnerability risk by 2.12$\times$~\cite{liu2026skillswild}: T1 and T2 skills are never granted script execution.

\textit{(3) Lifecycle trust evolution.} Deployed skills are subject to continuous runtime monitoring. Anomalous behavior (unexpected tool calls, permission boundary probes) triggers demotion or revocation. Conversely, skills with clean runtime histories can be promoted. This creates an incentive structure analogous to reputation systems in package management ecosystems.

The framework is deliberately architecture-aware: it maps directly to the three progressive disclosure levels identified in Section~\ref{sec:architecture}. Level~1 metadata is the only component exposed at T1; Level~2 instructions are accessible at T2 and above; Level~3 executable scripts require T3 or T4 trust. This correspondence ensures that governance decisions are grounded in the actual attack surface rather than applied uniformly.

A practical enterprise implementation could integrate these gates into the same lifecycle already used for internal packages. At intake, a registry service records provenance and runs G1--G2 before a skill is searchable by agents. For skills requesting scripts, file-system access, network access, or external MCP servers, a sandboxed staging environment runs G3 and compares observed behavior with the permission manifest in G4. At deployment, the runtime enforces the resulting tier by limiting which files, tools, network destinations, and approval scopes the agent may use while that skill is active. This is a governance proposal rather than an empirically validated system; its value is to make the trust assumptions explicit and testable.

\section{Open Challenges and Research Agenda}
\label{sec:challenges}

We identify seven open challenges that define the frontier of agent skills research:

\paragraph{Challenge 1: Cross-Platform Portability.} While Agent Skills has been released as an open standard, true cross-platform portability remains aspirational. Skills authored for Claude may implicitly depend on Claude-specific capabilities (code execution environment, tool signatures, model behaviors). Achieving genuine portability requires either (a) a universal skill runtime or (b) skill compilation targeting different agent platforms.

\paragraph{Challenge 2: Skill Selection at Scale.} Li~\cite{li2026singleagent} identified a phase transition in skill selection accuracy as library size grows. As enterprise skill libraries scale to hundreds or thousands of skills, the routing problem---determining which skill(s) to activate for a given query---becomes combinatorially challenging. Advanced Tool Use features~\cite{anthropic2025advancedtools} address this partially through Tool Search, but the fundamental scaling problem persists.

\paragraph{Challenge 3: Skill Composition and Orchestration.} Real-world tasks often require composing multiple skills. CUA-Skill's composition graphs~\cite{chen2026cuaskill} and Agentic Proposing's dynamic composition~\cite{jiao2026agenticproposing} offer initial solutions, but principled frameworks for multi-skill orchestration---including conflict resolution, resource sharing, and failure recovery---remain underdeveloped.

\paragraph{Challenge 4: Capability-Based Permission Models.} Current skill execution operates with implicit trust: once loaded, a skill can direct the agent to use any available tool. The security findings of Section~\ref{sec:security} demonstrate that this trust model is exploitable. A capability-based permission system, where each skill declares required permissions and the agent or user grants them explicitly, would significantly reduce the attack surface.

\paragraph{Challenge 5: Skill Verification and Testing.} Unlike software packages with unit tests and CI/CD pipelines, skills currently lack standardized testing frameworks. Anthropic's eval guidance~\cite{anthropic2025evals} provides principles but not skill-specific tooling. Automated skill verification---confirming that a skill does what it claims and nothing more---is an open technical problem that intersects AI safety and formal methods. Evaluation should include negative tests for hallucinated procedures, cascading execution failures, deadlocks between conflicting skills, silent permission escalation, and adversarial skill chaining.

\paragraph{Challenge 6: Continual Skill Learning Without Catastrophic Forgetting.} Shenfeld \etal{}~\cite{shenfeld2026selfdistillation} investigate whether pretrained LLMs can acquire new skills without degrading existing capabilities, finding that self-distillation offers a promising path. However, the interaction between dynamically loaded skills and the model's base capabilities---whether skills can inadvertently ``overwrite'' useful default behaviors---remains poorly understood.

\paragraph{Challenge 7: Evaluation Methodology.} Current benchmarks evaluate agents on task completion but rarely assess skill quality directly. Metrics for skill \textit{reusability} (does the skill generalize across tasks?), \textit{composability} (can it be combined with other skills?), and \textit{maintainability} (how robust is it to environmental changes?) are needed to evaluate skill ecosystems rather than individual agent runs.

\section{Discussion}
\label{sec:discussion}

The agent skills paradigm represents a shift from \textit{monolithic intelligence} to \textit{modular expertise}. This shift has practical consequences that extend beyond technical architecture. For organizations, skills offer a mechanism to encode institutional knowledge in a form that survives personnel turnover---a digital analogue of standard operating procedures. For the AI ecosystem, open skill standards create network effects: every skill contributed to the commons increases the value of the platform for all users.

The security challenges, however, are real and urgent. The 26.1\% vulnerability rate found by Liu \etal{}~\cite{liu2026skillswild} is not merely a technical metric; it reflects the fundamental tension between openness and safety that has characterized every major software ecosystem from package managers to app stores. Our proposed Trust and Lifecycle Governance Framework (Figure~\ref{fig:trust}) offers a principled path forward by decoupling trust decisions from binary accept/reject into graduated tiers that align permissions with provenance and verification depth. It also highlights non-malicious failure modes: a benign skill can hallucinate outdated instructions, trigger a cascade of tool failures, deadlock when composed with another skill, or overgeneralize an approval into a broader permission boundary. The skill ecosystem is currently in its ``pre-governance'' phase, and the decisions made in the coming months about verification pipelines, permission models, and trust hierarchies will shape its trajectory for years.

From a research perspective, the most promising directions lie at the intersection of skill acquisition and deployment. SAGE~\cite{wang2025sage} and SEAgent~\cite{sun2025seagent} demonstrate that agents can learn skills through experience, but these learned skills are model-internal---they cannot be inspected, shared, or governed in the way that human-authored SKILL.md files can. Bridging this gap---enabling agents to not only learn skills but externalize them as portable, auditable artifacts---would unify the acquisition and deployment paradigms.

\section{Conclusion}
\label{sec:conclusion}

Agent skills have emerged as a foundational abstraction for the next generation of LLM-based agents. By packaging procedural expertise as composable, portable, and dynamically loadable modules, skills resolve the tension between general-purpose models and specialized task requirements. The rapid ecosystem growth---from Anthropic's October 2025 launch to open standardization in December 2025, with adoption across multiple frontier model providers---confirms that the industry has converged on this abstraction.

This survey has traced the skill paradigm across four dimensions: architectural foundations, acquisition methods, deployment settings with emphasis on the CUA stack, and security. The field's progress has been remarkable: benchmark results that seemed aspirational a year ago are now routine, and new challenges---security governance, cross-platform portability, skill composition---have emerged to define the next frontier.

The path forward requires simultaneous advances in multiple directions: more principled skill learning algorithms that produce inspectable artifacts, robust permission models that maintain usability, evaluation frameworks that assess skill quality rather than merely task completion, and governance structures that balance openness with safety. The agent skills paradigm is young, but its trajectory suggests it will be central to how humanity collaborates with AI systems in the years ahead.


\bibliographystyle{ACM-Reference-Format}
\bibliography{references}

\end{document}